\documentclass[9pt]{article}

\usepackage{amssymb} 
\usepackage{graphicx} \graphicspath{ {/home/nic/Dropbox/LorenzWithMemoryPaper/images/} }

\usepackage{graphicx} \graphicspath{ {images/} }

\usepackage{subfigure} 

\usepackage{appendix}

\usepackage{caption}

\usepackage{mathtools}   
\usepackage{cases}

\usepackage{sectsty}
\usepackage{datetime}

\sectionfont{\normalsize}
\subsectionfont{\normalsize}

\begin{document}

\title{\large \textbf{A model for dynamical systems with strange attractors}} 
\author{{\normalsize Nicola Romanazzi$^{\footnote{Email address: nrom@upenn.edu}}$ } \\
 {\small\it Philadelphia, PA 19104, USA}} 

\date{November 29, 2024}
\maketitle

\begin{abstract}

We derive a system with one degree of freedom that models a class of dynamical systems with strange attractors in three dimensions. This system retains all the characteristics of chaotic attractors and is expressed by a second-order integro-differential equation which mimics a spring-like problem. We  determine the potential energy, the rate of change of the kinetic energy of this system, and show that is self-oscillating.
 
\end{abstract}

\section{Introduction}

In this paper, we probe into the nature of a class of dynamical 
systems with strange attractors 
in three dimensions by transforming their classical equations of motion, 
specified by a set of first order differential equations (ODEs), 
into systems with one degree of freedom.
In particular, we study the link between energy and self-sustainability 
of oscillation (or self-oscillation). 
Specifically, we show that self-sustainability is manifested when 
the rate of change of energy is greater than zero,
for motion in a chaotic regime, and zero for limit cycles.
We carry out this study for two archetypal systems that fall
into this class: the R\"ossler system and the Lorenz system. 
We transform the first order ODEs that define the R\"ossler system \cite{rossler76},
  \begin{subequations}\label{eq:rossler_first_order}
    \begin{numcases}{}
      \dot{x} = -y-z \label{eq:rossler_first_order1} \\
      \dot{y} = x+ay \label{eq:rossler_first_order2} \\  
      \dot{z} = -cz+xz+b, \label{eq:rossler_first_order3}
    \end{numcases}
  \end{subequations}
and the Lorenz system \cite{lorenz63},
  \begin{subequations}\label{eq:lorenz-ode}
    \begin{numcases}{}
       \dot{x} = -\sigma(x-y)\label{eq:lorenz-ode1} \\  
       \dot{y} = rx-y-xz \label{eq:lorenz-ode2} \\  
       \dot{z} = -bz+xy, \label{eq:lorenz-ode3} 
    \end{numcases}
  \end{subequations}
into two respective second order integro-differential equations (IDE).
Through this transformation, we re-frame the description of their motion from kinematics to 
dynamics \cite{landau,festa2002}. The models that emerge from this transformation reduce 
the study of these fairly complex chaotic systems in three
dimensions into familiar one dimensional spring like models. 

This paper is organized as follows. In section 2.1, the procedure for the derivation of a second order IDE
for the R\"ossler system is discussed. In section 2.2, the structure of the IDE and its physical
interpretation is discussed. In section 2.3, 
we compute the time rate of change of kinetic energy, and show that
that the R\"ossler system is self-oscillating. In section 2.4 we discuss 
the potential energy of the R\"ossler system.
In section 3.1, we show the derivation of the Lorenz system's IDE. In section 3.2, 
we discuss the time rate of change of 
the energy of the Lorenz system. 
In section 3.3, we compute
the potential energy of the Lorenz system. In Section 3.4, we 
draw a parallel between the Duffing system and 
Lorenz system.  
In Section 4, we present some discussions about this model 
and we conclude in Section 5.

\section{A second order IDE for the R\"{o}ssler system}
\label{sec:rossler}

In this section, we discuss the derivation of the R\"ossler IDE. 
We also determine analytically the 
 equation of the mechanical potential, and
carry out analytical and numerical calculation of the time
rate of change of kinetic energy.

\subsection{The equation of motion for the R\"{o}ssler system}
\label{subsec:rossler1}

The derivation of a second-order IDE as an equation of motion for the 
R\"{o}ssler system is entirely analytic, 
and revolves around the use of variable substitutions.
This derivation separates in two main tasks: first, it expresses the nonlinear term $xz$, 
shown in Equ:~\eqref{eq:rossler_first_order3}, as the sum of 
linear terms in $y$ and its derivatives; 
second, it relies on an analytic solution of a first order differential equation~\eqref{eq:rossler_first_order3} 
that emerges in the derivation.
We explain further.

Consider Equ:~\eqref{eq:rossler_first_order1},
and rewrite it as $z=-\dot{x}-y$.
Multiplying this equation by $x$ 
\begin{equation} \label{eq:ross_non_linear_term}
xz= -x\dot{x} -xy,
\end{equation}
and using a relation
between the variables $x$ and $y$ from Equ: \eqref{eq:rossler_first_order2}, yields 
\begin{equation*} 
x=\dot{y}-ay
\end{equation*}
and its first derivative.
\begin{equation*} 
\dot{x}=\ddot{y}-a\dot{y}.
\end{equation*}
Through variable substitutions of $x$ and $\dot{x}$ in the right-hand side of 
Equ: \eqref{eq:ross_non_linear_term}, we find an expression for the 
nonlinear term $xz$ in terms of $y$ and its derivatives,
\begin{eqnarray}
xz = -\dot{y}\ddot{y} +\dot{y}^2 + a y\ddot{y} - a^2 y \dot{y} - y \dot{y} + a y^2,
\label{eq:rossler_nonlinear_term2} 
\end{eqnarray}
which leads us to write Equ: \eqref{eq:rossler_first_order3} as 
\begin{eqnarray}
\dot{z}+ cz = b -\dot{y}\ddot{y} +\dot{y}^2 + a y\ddot{y} - a^2 y \dot{y} - y \dot{y} + a y^2.
\label{eq:rossler_z_first_order} 
\end{eqnarray}
This differential equation is fully solvable, and the solution explicitely derived 
as follows.  Define a function $M(t)$ as
\begin{equation} \label{eq:mt}
M(t)=\frac{1}{2} \dot{y}^2 - ay\dot{y} + \frac{a^2}{2} y^2 - \frac{b}{c} + z,
\end{equation}  
and its derivative $\dot{M}(t)$
\begin{equation*} 
\dot{M}(t)= \dot{y}\ddot{y} -a \dot{y}^2 - a y\ddot{y} + a^2y\dot{y}+\dot{z}.
\end{equation*}  
The differential equation, Equ:~\refeq{eq:rossler_z_first_order}, can be re-written as
\begin{eqnarray}
\dot{M}(t)+ cM(t) = S(t),
\label{eq:rossler_general_first_order} 
\end{eqnarray}
where $S(t)$ is a function in terms of $y$ and its derivatives \cite{schultz1967state}.
To find an explicit expression for $S(t)$, we first substitute the
functions $M(t)$ and $\dot{M}(t)$, given in terms of $y$, $\dot{y}$, and $\ddot{y}$, into 
Equ: \eqref{eq:rossler_general_first_order}:
\begin{eqnarray}
\dot{z}+cz=S(t)+b-\dot{y}\ddot{y}+a\dot{y}^2+ay\ddot{y}-a^2 y\dot{y}-\frac{c}{2}\dot{y}^2+ac y\dot{y}-\frac{a^2}{2}cy^2.
\label{eq:rossler_s} 
\end{eqnarray}
Equating Equ: \eqref{eq:rossler_z_first_order} and Equ: \eqref{eq:rossler_s} yields 
 $S(t)$ as
\begin{equation} \label{eq:st}
S(t)=\frac{c}{2}\dot{y}^2-(ac+1)y\dot{y}+a(1+\frac{ac}{2})y^2.
\end{equation} 
Eq: \eqref{eq:rossler_general_first_order} admits a solution of the form \cite{schultz1967state} 
\begin{eqnarray}  
M(t) = M(0) e^{-ct} + \int_{0}^t e^{-c\tau} S(t-\tau) d\tau.
\end{eqnarray}
This type of differential equations have been studied in the context 
of feedback, and is widely used in electrical engineering \cite{schultz1967state}.
Since we are interested in the steady state solution of the equation, 
we discard the transient solution
by letting $t \rightarrow \infty$. In this way, $M(0) e^{-ct}$ converges to zero. Hence
\begin{eqnarray} \label{eq:rossler_steady_state}
M(t) = \int_{0}^t e^{-c\tau} S(t-\tau) d\tau.
\end{eqnarray}
By substituting the functions $M(t)$, Equ:~\refeq{eq:mt}, and  $S(t)$, Equ:~\ref{eq:st}, 
in Equ: \eqref{eq:rossler_steady_state}, 
we obtain an equation of the form 
\begin{equation*} 
z+\dot{y}^2-ay\dot{y}+\frac{a^2}{2}y^2-\frac{b}{c}=\int_{0}^t e^{-c\tau} S(t-\tau) d\tau.
\end{equation*} 
Note that the $z$ variable still appears in this equation. It can be eliminated by 
using the relation
 $z=-\ddot{y}+a\dot{y}-y$, which derives from Equ: \eqref{eq:rossler_first_order1} and 
Equ: \eqref{eq:rossler_first_order2}. After careful substitution of the $z$ variable, 
the form of a second order IDE  in the variable $y$ and its derivatives  is derived as
\begin{eqnarray}
\ddot{y}-(a-ay+\dot{y})\dot{y}+y+\frac{a^2}{2}y^2+\frac{b}{c}+\int_{0}^t e^{-c\tau} S(t-\tau) d\tau =  0
\label{eq:rossler_final_second_order} 
\end{eqnarray}
where $S(t-\tau)=[c\dot{y}-(ac+1)y]\dot{y}+(a+\frac{ac}{2})y^2$. 

A close look at this IDE allows to recognize frictional and elastic forces. 
We rearrange the terms given by the 
convolution integrals in the equation, also known as heredity, retarded reaction, or memory terms \cite{memory0,memory1,memory2}, and 
identify them as contribution terms to the frictional force and elastic force \cite{langevin1,volterra}. 
We then write the R\"ossler IDE in its final form, after a final substitution of $y$ 
and its derivatives by the generic variable $q$ and its derivatives ($y=q$, $\dot{y}=\dot{q}$, $\ddot{y}=\ddot{q}$): 
\begin{equation} \label{eq:rossler-second-order-final}
\begin{split}
&\ddot{q} + [a(q-1)-\frac{1}{2}\dot{q}]\dot{q}+\int_{0}^t e^{-c\tau} [(ac+1) q(t-\tau)-\frac{c}{2}\dot{q}(t-\tau)]\dot{q}(t-\tau) d\tau \\ &+ q - \frac{a^2}{2} [q^2 
+ (c+\frac{2}{a}) \int_{0}^t e^{-c\tau}   q^2(t-\tau)  d\tau] = - \frac{b}{c}.\\
\end{split}
\end{equation}
We recognized the damping force of the system as 
$- [a(q-1)-\frac{1}{2}\dot{q}]\dot{q}-\int_{0}^t e^{-c\tau} [(ac+1) q(t-\tau)-\frac{c}{2}\dot{q} (t-\tau)]\dot{q}(t-\tau) d\tau$, 
the elastic force as $- q + \frac{a^2}{2} [q^2 + (c+\frac{2}{a}) \int_{0}^t e^{-c\tau}   q^2(t-\tau)  d\tau]$, along with a constant force given by $- \frac{b}{c}$.

\begin{figure}
\centering     
\subfigure[Period-one limit cycle: control parameters $a=0.2,b=2.0, c=4.0$]
   {\label{fig:rossler_one_limit_cycle}
    \includegraphics[width=50mm]{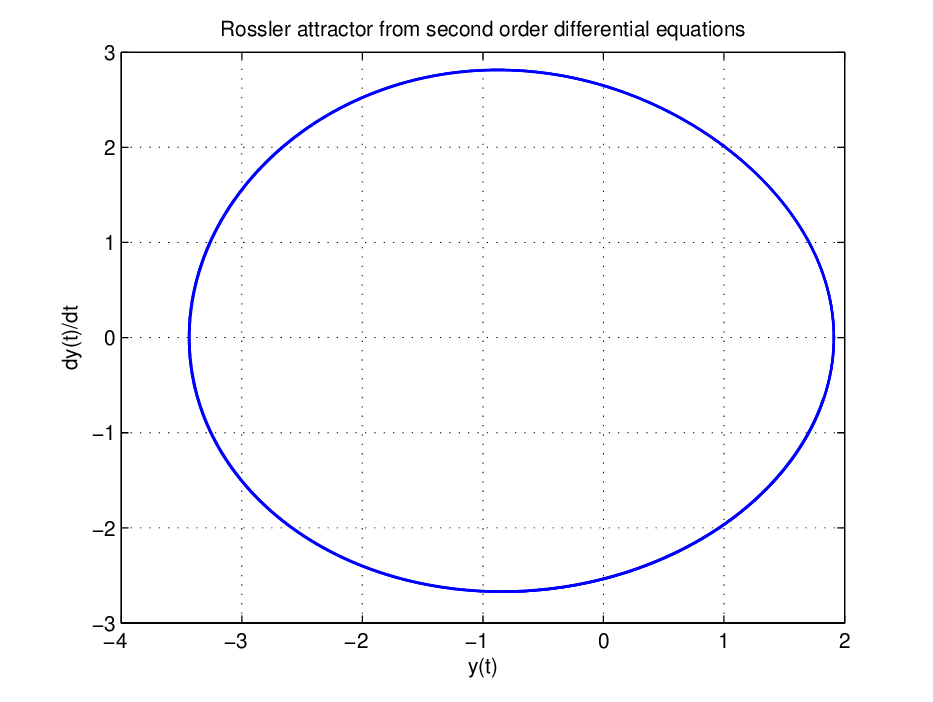}}
\subfigure[Period-two limit cycle: control parameters $a=0.35,b=2.0, c=4.0$]
   {\label{fig:rossler_two_limit_cycle}
   \includegraphics[width=50mm]{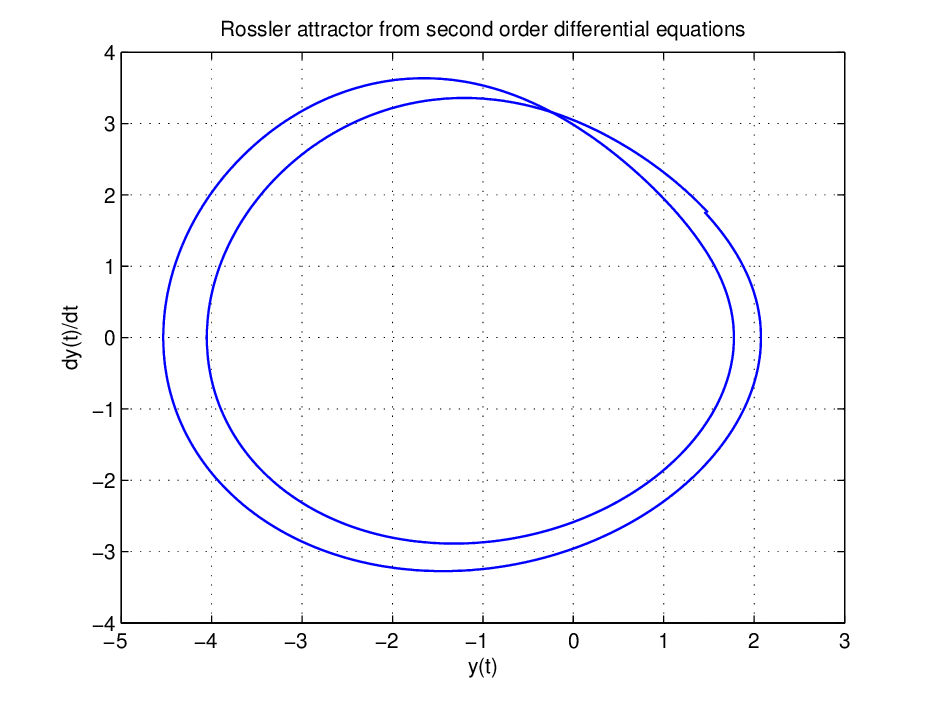}}
\caption{  Numerical solution of Eq.~(\ref{eq:rossler_final_second_order}) for limit cycles} 
\label{fig:rossler_chaos} 
\end{figure}

   



The solutions of this second order IDE, Equ: \eqref{eq:rossler-second-order-final}, are 
equivalent to the solutions of the set 
of three first order ODEs, Equ: \eqref{eq:rossler_first_order}, that define the R\"ossler system
in three dimensions.  This follows from the
analytic and exact nature of the IDE. 
For completeness, we numerically solve the IDE,  using 
the classical Runge-Kutta forth-order method.

In particular, we carry out the calculation for a number of sets of
control parameters, for which the solutions of the ODEs are well known and illustrative: 
a limit cycle of period one, Fig:~\ref{fig:rossler_one_limit_cycle}, a limit cycle of period two, Fig:~\ref{fig:rossler_two_limit_cycle}; and the chaotic attractor in the
 chaotic regime, Fig:~\ref{fig:rossler_chaos}. 
\begin{figure}
\centering     
\subfigure[R\"ossler system in the chaotic regime, control parameters $a=0.432,b=2.0, c=4.0$]
   {\label{fig:rossler_chaos_2d}\includegraphics[width=50mm]{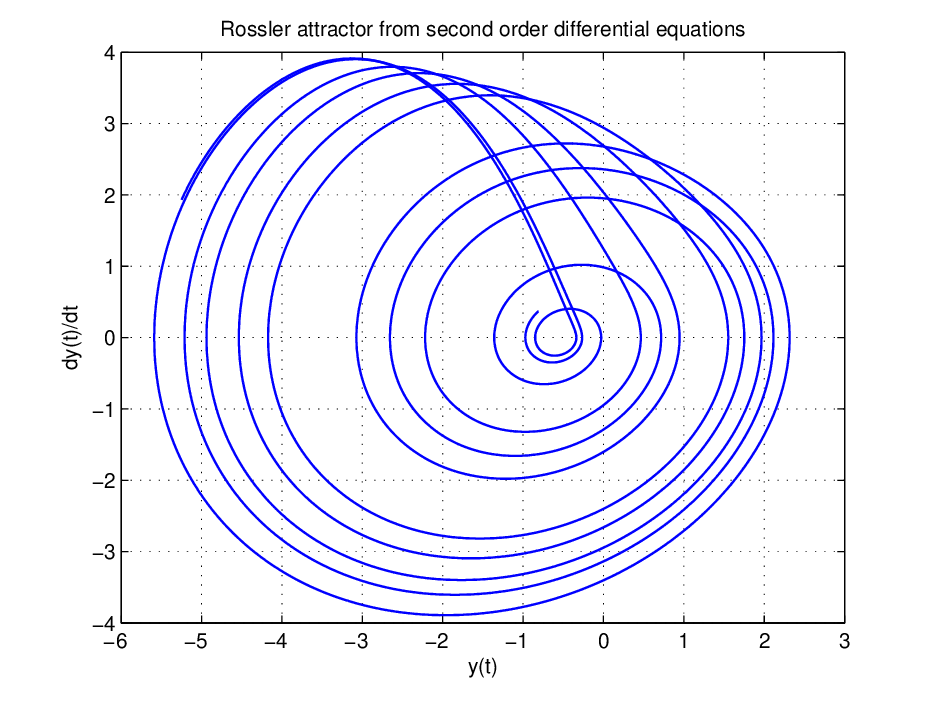}}
\subfigure[Same as (a) in three dimensions]
   {\label{fig:rossler_chaos_3d}\includegraphics[width=50mm]{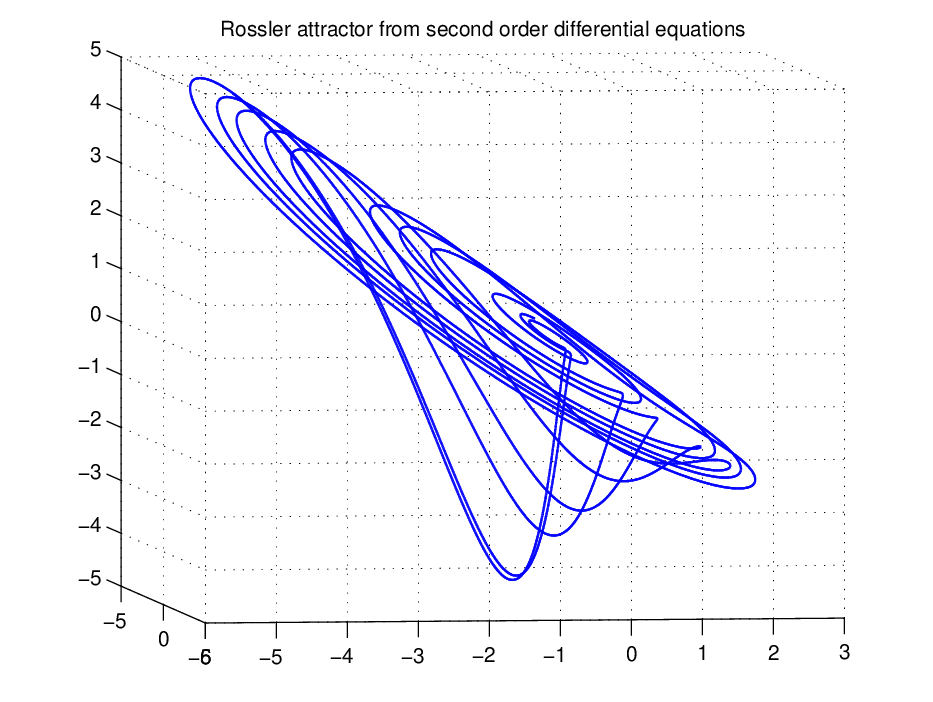}}
\caption{ A numerical solution of Eq.~(\ref{eq:rossler_final_second_order}) in the 
chaotic regime} \label{fig:rossler_chaos} 
\end{figure}
As we can see from the plots of the numerical solutions of the IDE conforms with the known solutions the ODEs.

\subsection{What does the structure of the IDE say about the physical model?}
\label{subsec:rossler2}

The general form of the R\"ossler IDE suggests some basic properties of the physical 
model it represents.
One feature, so revealed from the IDE, is that the model it supports can be described as the motion of a unit mass in 
a potential, whose shape can be deduced from the form of the IDE.

Let us write the R\"ossler IDE in a shorthand notation, 
\begin{equation} \label{eq:rossler_anharmonic_1}
\ddot{q}+h(\dot{q},q)+\beta q + \alpha M(q^n) + F = 0
\end{equation}
where
$h(\dot{q},q)=[a(q-1)-\frac{1}{2}\dot{q}]\dot{q}-\int_{0}^t e^{-c\tau} [(ac+1) q(t-\tau)-\frac{c}{2}\dot{q} (t-\tau)]\dot{q}(t-\tau) d\tau$;
and $M(q^2)=[q^2 + (c+\frac{2}{a}) \int_{0}^t e^{-c\tau}   q^2(t-\tau)  d\tau]$.
The coefficients $\beta=1$, $\alpha=-\frac{a^2}{2}$, and $n=2$ determine the R\"ossler system.
We recognize Equ: \eqref{eq:rossler_anharmonic_1} as the equation of a 
dissipative anharmonic oscillator; in particular, the R\"ossler system can be modelled by a dissipative
softening spring. The system evolves in a one-well mechanical potential.  

We refer the reader to Appendix A where we examine three possible IDE configurations defined through
the values of the coefficients
$\beta$, $\alpha$ and $n$, with $n$ being the power in the correction 
term to the elastic force in the IDE.   
These three types of systems correspond to a hardening spring system in a one-potential well, this
being the case of the Duffing oscillator;  a hardening spring in a two-well potential, this being the case of the
Lorenz system; and a softening spring in a one-well potential, modeling 
the R\"ossler system.

\subsection{Time rate of change of the kinetic energy}
\label{subsec:rossler3}

The R\"ossler system is known to be dissipative, though maintaining oscillations without an external driving force.
This is apparent from the solutions of the ODEs and reinforced from the 
form of the R\"ossler IDE.
In this section, we show that 
the self-oscillatory motion in the R\"ossler system is caused by the existence of a positive 
rate of change in energy, per cycle of oscillation.

\begin{figure}[t!]
\centering
\includegraphics[width=6cm]{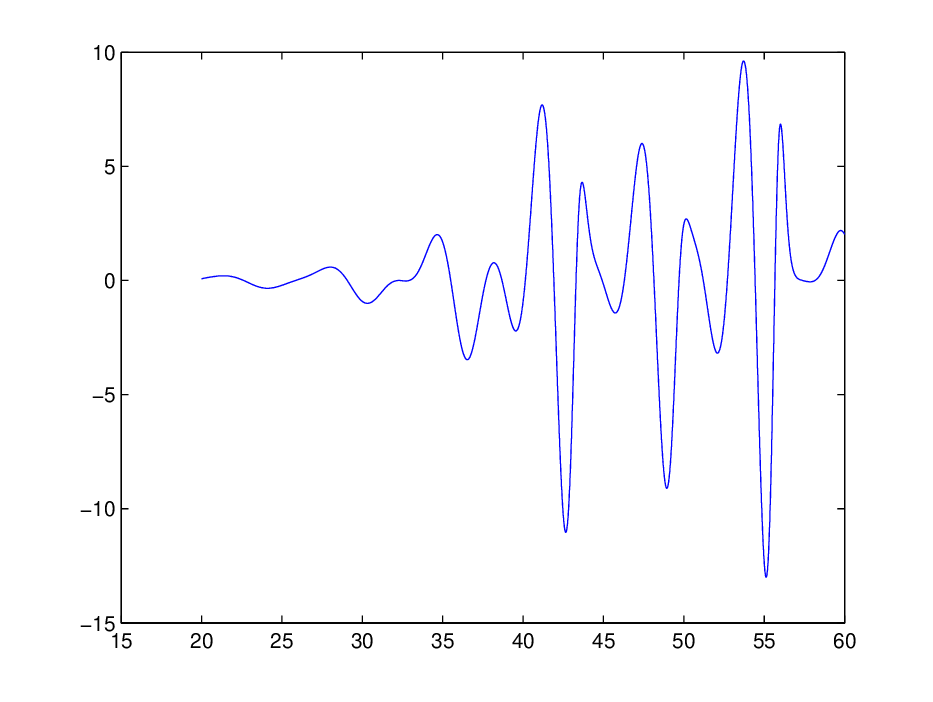}
\caption{R\"ossler time rate of change of kinetic energy for 
control parameters (a=0.432,b=2.0, c=4.0)--chaotic regime.}
\label{fig:time_rate_energy_K}
\end{figure}

Rearranging the terms of Equ: \eqref{eq:rossler_anharmonic_1} and multiplying 
both sides of the 
equation  by $\dot{q}$ yields
\begin{eqnarray}
\ddot{q} \dot{q} &=& -F \dot{q} - h(q,\dot{q}) \dot{q} - \beta q \dot{q} + \alpha M(q^2) \dot{q}.
\label{eq:rossler_work_1}
\end{eqnarray}
In this way, 
we obtain the rate of change of kinetic energy $\frac{dE_K}{dt}$ \cite{classicalmechanics}, where $E_K$ is easily determined
by integrating the left side of Equ:\eqref{eq:rossler_work_1},
$E_K = \frac{1}{2} \dot{q}^2$.
In determining the rate of change of kinetic energy, we took into account all the forces in the system.
We compute $\frac{dE_K}{dt}$ numerically using different control parameters;
$\frac{dE_K}{dt}$ fluctuates between positive and negative values, showing, however, a net energy gain per cycle,  
$\frac{dE_K}{dt}>0$, implying that the system pumps energy in, as in the case of the system with 
control parameters $a=0.432,b=2.0, c=4.0$, Fig: \ref{fig:time_rate_energy_K}.
The self-sustainable motion emerges from a careful balancing of the forces which act on the system
in such a way to enable
enough energy to be pumped into the system. Most significantly, this behavior of the system seems
to be governed by its memory terms. This also causes 
the irregular oscillation of an observable (q)
when the system runs in the chaotic regime, as 
the amount of energy pumped in is also irregular cycle-by-cycle. We show in Appendix B that the existence of a memory
term is a necessary condition that determines a rate of change of kinetic energy greater than zero.

\begin{figure}[t!]
\centering
\includegraphics[width=6cm]{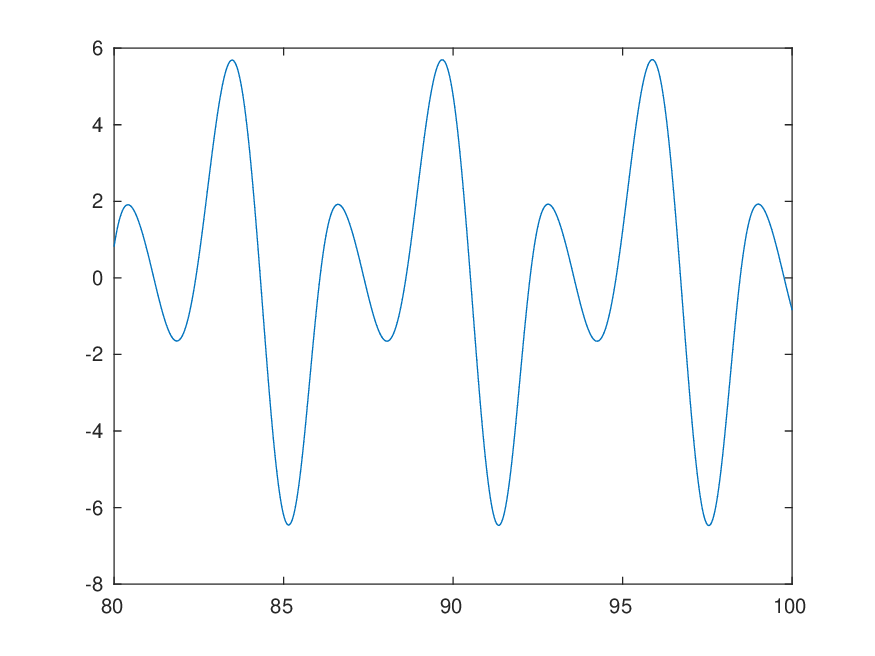}
\caption{R\"ossler time rate of change of kinetic energy for 
limit cycle, control parameters $a=0.2,b=2.0, c=4.0$.}
\label{fig:limitycycleenergy}
\end{figure}

In the case of limit cycles, the system is still dissipative, as shown by the equation of motion; 
however, the regular pattern of oscillation of the motion comes from
a rate of exchange of energy equals to zero, $\frac{dE_K}{dt}=0$, per cycle. 
In other words, over one cycle, equal amounts
of energy are pumped in and removed from the system, making the system 
still self-sustainable, but regular in its oscillation, Fig:\ref{fig:limitycycleenergy}.

In the course of this study, we have observed that, contrary to intuition,  the dissipative force of
the R\"ossler system produces, what in the literature is known as, negative dissipation \cite{jenkins}---negative 
dissipation being a dissipative force that acts in such a way to pump energy into a system.
\begin{figure}[t!]
\centering
\includegraphics[width=6cm]{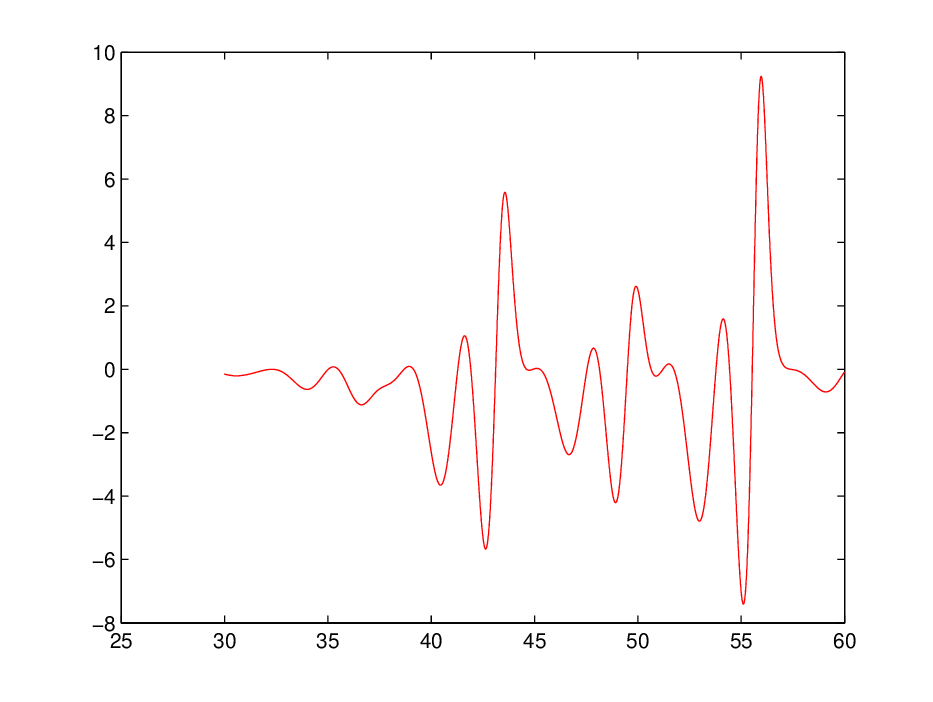}
\caption{R\"ossler time rate of change of dissipation for 
control parameters (a=0.432,b=2.0, c=4.0)--chaotic regime. The dissipation alternates between positive and negative
dissipation}
\label{fig:negative_dissipation}
\end{figure}
In Figure \ref{fig:negative_dissipation}, we show the energy rate of change of the dissipation versus time, where
clearly, the dissipation alternates between positive dissipation and negative dissipation. We use the
term positive dissipation in juxtaposition to negative dissipation---positive dissipation being to be interpreted in
the classical sense, as a force that takes energy away from a system.
In the R\"ossler system, dissipation acts as a force that can pump energy into the system which
seems one of the significant factors 
that determines self-oscillation.

\subsection{Potential energy}
\label{subsec:rossler4}

Given that the dynamics of the equation of motion is expressed by an IDE,  
from Equ: \refeq{eq:rossler_anharmonic_1}, we can derive the mechanical potential $U(q)$ of the system. 
The potential simply results from the integration of the elastic force $\beta q - \alpha M(q^2)$: here, $\beta q - \alpha M(q^2) = q-\frac{a^2}{2}
[q^2 + (c+\frac{2}{a}) \int_{0}^t e^{-c\tau}   q^2(t-\tau)  d\tau]$. Therefore, the potential is
\begin{equation}
U(q,t) = \frac{1}{2} q^2 - \frac{a^2}{6} q^3  + (c+\frac{2}{a}) \int \int_{0}^t e^{-c\tau} q^2(t-\tau)d\tau dq.
\label{eq:rossler_potential}
\end{equation}
This potential can be thought of as being defined by the sum of two terms: 
one term that depends on the variable $q$, $U(q)=\frac{1}{2} q^2 - \frac{a^2}{6} q^3$, 
and a second
term that depends on $q$ and $t$, $U(q,t)=(c+\frac{2}{a}) \int \int_{0}^t e^{-c\tau} q^2(t-\tau)d\tau dq$.
Because of the dependency on time $t$, the potential can be shown 
to be formed by a family of curves; each curve 
contributes to a 
distortion of the potential curve that varies cycle by cycle. 

\begin{figure}[t!]
\centering
\includegraphics[width=6cm]{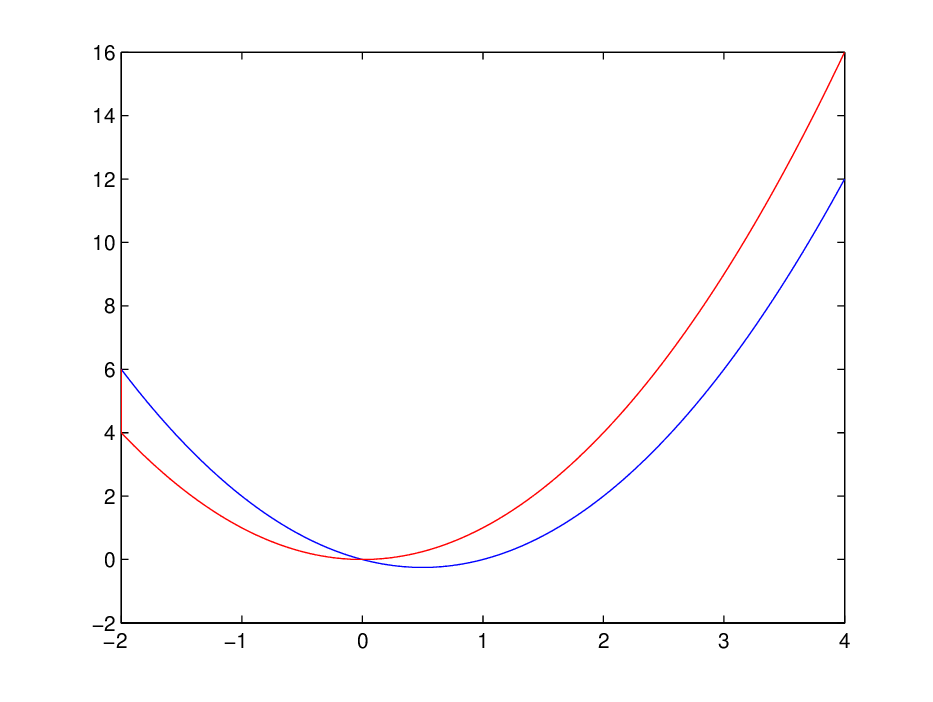}
\caption{Plot of the function $y=x^2-x$ (blue curve), and $y=x^2-\int e^{-bt} x(t-\tau) d\tau$ (red curve), with b=0.4.}
\label{fig:memory_effect}
\end{figure}

To acquire an appreciation of how a memory term might affect a function, 
consider
a simple function, for example $y=x^2-x$, and a related 
function $y=x^2-\int e^{-bt} x(t-\tau) d\tau$
obtained by
substituting the variable $x$ with the memory term $\int e^{-bt} x(t-\tau) d\tau$.
The memory term accounts for the entire 
history of the system.
We plot the two functions in Fig: \ref{fig:memory_effect},
which 
shows an evident distortion of the curve caused by the action of the memory term.   
This type of distortion is naturally present in the potential. 
\begin{figure}
\centering     
   {\label{fig:rossler_chaos_2d}\includegraphics[width=60mm]{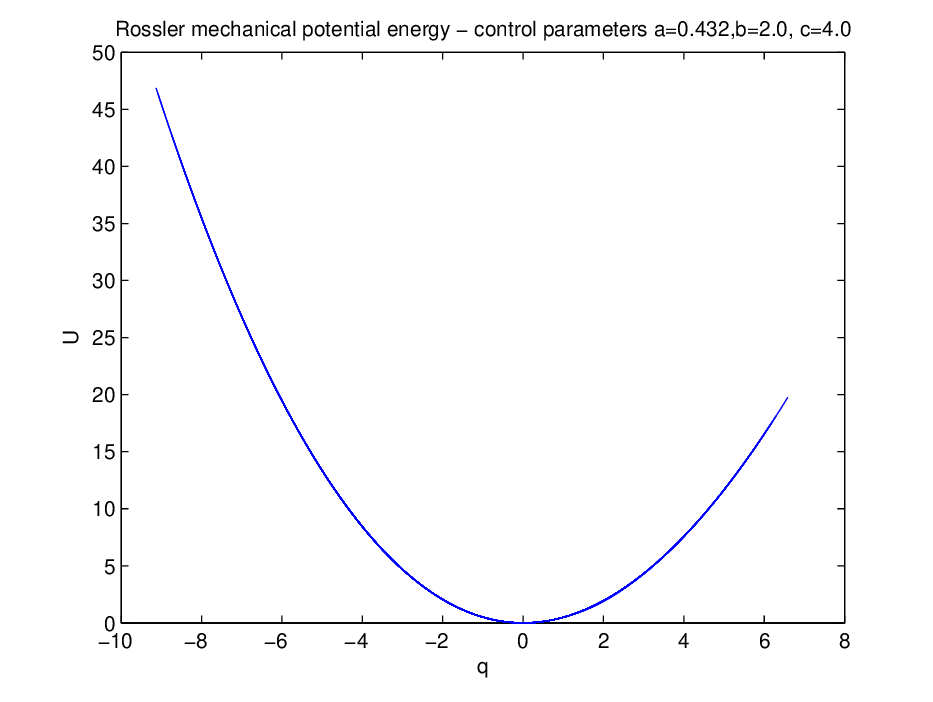}}   
\caption{Plot of R\"ossler potential, Equ:~\eqref{eq:rossler_apprx_potential}}
 \label{fig:rossler_potential_energy} 
\end{figure}

In plotting the potential curve of the R\"ossler system, it is desirable to 
reduce the family of curves
that emerge from the presence of the memory term,
to one curve. We show in Fig:~\ref{fig:rossler_potential_energy} such a curve obtained
 by replacing 
the kernel $e^{-bt}$, in the memory term of the potential $U(q,t)$, Equ: \refeq{eq:rossler_potential}, 
with a delta function: $U(q) = \frac{1}{2} q^2 - \frac{a^2}{6} q^3  + (c+\frac{2}{a}) \int \int_{0}^t \delta(t-\tau) q^2(\tau)d\tau dq$.
This allows to derive a potential curve that retains the general 
characteristics of the actual potential.  In this way, the integral 
above is in close form, and $U(q)$ expresses an approximated potential of the form
\begin{equation}
U(q) = \frac{1}{2} q^2 - \frac{a^2}{6} q^3  + (c+\frac{2}{a}) \frac{1}{3} q^3.
\label{eq:rossler_apprx_potential}
\end{equation}
We calculated the potential $U(q)$ numerically  for three sets of control parameters; Fig:~\ref{fig:rossler_potential_energy} shows the potential of a period one limit cycle ($a=0.2,b=2.0, c=4.0$),
for a period two  limit cycle ($a=0.35,b=2.0, c=4.0$),
and for the system in the chaotic regime ($a=0.432,b=2.0, c=4.0$). The three curves overlap,
although, we note that each curve is determined by branches of different lengths.

\section{A second order IDE for the Lorenz system}
\label{sec:lorenz}

In this section we discuss the derivation of the Lorenz IDE. 
We derive the equation of motion of the Lorenz system, 
its mechanical potential, and the rate of change of energy.

\subsection{The equation of motion for the Lorenz system}
\label{subsec:lorenz1}

\begin{figure}[]
\begin{subfigure}[]
   {\includegraphics[width=6cm]{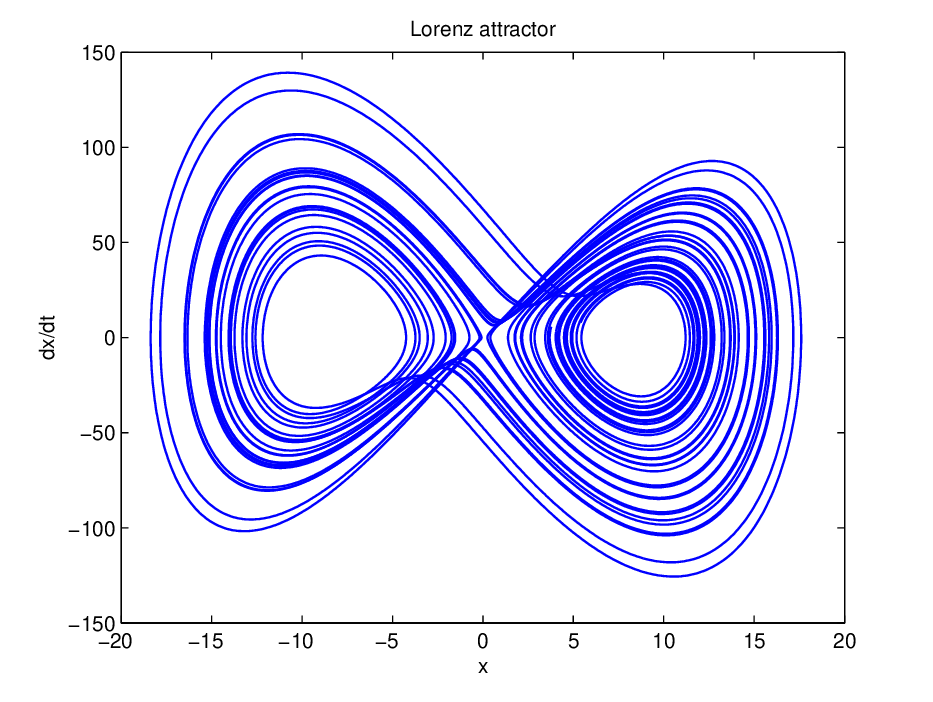}} 
\end{subfigure}
\begin{subfigure}[]{\includegraphics[width=6cm]{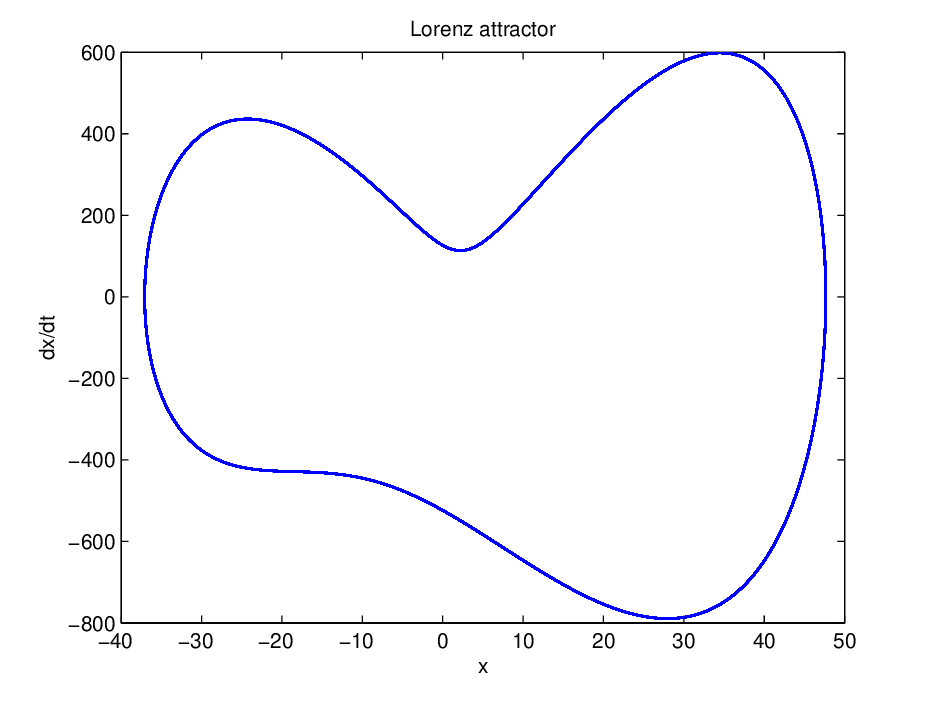}}
\end{subfigure}
 \caption{ Numerical Solution \cite{brunner1987} of Eq.~(\ref{eq:lorenz-in-x}). 
(a) The parameters values, $\sigma = 10$, $b = 8/3$, and $r = 28$, are those chosen 
by Lorenz to form his classical \textit{Butterfly Chaotic Attractor}. 
(b) A Limit Cycle with parameters $\sigma = 10$, $b = 8/3$, and $r = 237$ is also shown.} 
\label{fig:lorenzwithouttransient} 
\end{figure}

The derivation of the Lorenz second order IDE follows closely, but not identically, the 
derivation the R\"ossler IDE. 
The idea is to recombine the variables $\dot{x}$, $\dot{y}$,
and $\dot{z}$, into one variable equation \cite{festa2002}. Here, we express $y$ in
terms of $x$ and $\dot{x}$:  $y = \frac{1}{\sigma} (\dot{x} + \sigma
x)$, followed by the $y$ substitution into $\dot{z} = -bz+xy$.
This yields the ODE 
\begin{equation}
\dot{z} + b z = \frac{1}{2 \sigma} (2 x \dot{x} + 2 \sigma x^2). 
\label{eq:lorenz-z}
\end{equation} 
Equation~(\ref{eq:lorenz-z}) is of the form $\dot{M}(t) + b M(t) =
S(t)$, which admits a solution of the form \cite{schultz1967state}
\begin{equation}\label{eq:lorenz-z-solution}
z(t) = -e^{-bt} z(0) + \frac{1}{2 \sigma} x^2(t) - (\frac{b}{2\sigma} - 1)
\int_{0}^t e^{-b\tau} x^2(t-\tau) d\tau 
\end{equation}
Since we are only interested in the steady state solution, 
we formally eliminate the transient solution from the above equation 
by taking the limit
$t \rightarrow \infty$, resulting in 
\begin{equation} \label{eq:lorenz-z-solution-no-z-transient}
z(t) = \frac{1}{2 \sigma} x^2(t)  - (\frac{b}{2\sigma} - 1)
\int_{0}^t e^{-b\tau} x^2(t-\tau) d\tau 
\end{equation} 
Finally,  substituting $y$, $\dot{y}$, and $z$ 
(from Eq.~(\ref{eq:lorenz-z-solution-no-z-transient})) into 
$\dot{y} = rx-y-xz$ (Eq.~(\ref{eq:lorenz-ode})), yields a second order IDE 
in terms of $x(t)$,
\begin{equation} \label{eq:lorenz-in-x}
\ddot{x} + (\sigma+1) \dot{x} - \sigma(r-1)x + 
\frac{1}{2} x^3 + 
(1-\frac{b}{2\sigma})x \int_{0}^t e^{-b\tau} x^2(t-\tau) d\tau = 0 
\end{equation}
Note that no approximation was made in deriving 
Eq.~(\ref{eq:lorenz-in-x}), implying
a complete equivalence between the solutions of Eq.~(\ref{eq:lorenz-in-x}) 
and Eq.~(\ref{eq:lorenz-ode}).
For instance, Fig. \ref{fig:lorenzwithouttransient}(a,b) shows
the typical butterfly chaotic attractor and a limit cycle. These solutions were obtained by
numerically solving Eq.~(\ref{eq:lorenz-in-x}).
When expressed in the form of Equ:~(\ref{eq:lorenz-in-x}), the Lorenz 
system clearly appears as an oscillator
of unit mass, with $(\sigma+1)$ being a damping term, and   
an elastic force $(1-\frac{b}{2\sigma})x \int_{0}^t
e^{-b\tau} x^2(t-\tau) d\tau$.
We write the Lorenz equation of motion in its final form by substituting the variable $x$ and its derivatives
with the variable $q$ and its derivatives,
\begin{equation}
\ddot{q} + (\sigma+1) \dot{q} - \sigma(r-1)q + 
\frac{1}{2} q^3 + 
\sigma(1-\frac{b}{2\sigma})q \int_{0}^t e^{-b\tau} q^2(t-\tau) d\tau = 0.
\end{equation}

\subsection{Time rate of change of the kinetic energy}
\label{subsec:lorenz2}

In this section, we compute 
the rate of change of the kinetic energy.
We write the Lorenz second order equation of motion in shorthand notation
\begin{eqnarray}
\ddot{q} + h(\dot{q}) - \beta q  + \alpha M(q,q^3) &=& 0, 
\label{eq:hardening_of_oscillators}
\end{eqnarray}
where the $h(\dot{q}) = (\sigma+1)\dot{q}$, $\beta = \sigma(r-1)$, $\alpha=+1$, and $M(q,q^3)=\frac{1}{2} x^3 +\frac{b}{2\sigma}-1)x \int_{0}^t e^{-b\tau} x^2(t-\tau) d\tau$.
By rearranging the terms of Equ:\ref{eq:hardening_of_oscillators} as
\begin{equation} \label{eq:lorenz_work_2}
\ddot{q}  = - h(q,\dot{q}) + \beta q  - \alpha M(q^3)
\end{equation}
and multiplying by $\dot{q}$, we obtain
\begin{equation} \label{eq:lorenz_work_3}
\ddot{q} \dot{q} = - h(q,\dot{q}) \dot{q} - \beta q \dot{q} + \alpha M(q^3) \dot{q}.
\end{equation}
Integrating the left hand side of Equ: \refeq{eq:lorenz_work_3}, we derive 
an expression for the rate of change of the kinetic energy,
\begin{equation}\label{eq:lorenz_work_4}
\frac{dE_k}{dt}  = - h(q,\dot{q}) \dot{q} - \beta q \dot{q} + \alpha M(q^3) \dot{q}
\end{equation}
where $E_k=\frac{1}{2}\dot{q}^2$. 
\begin{figure}
\centering     
   \includegraphics[width=60mm]{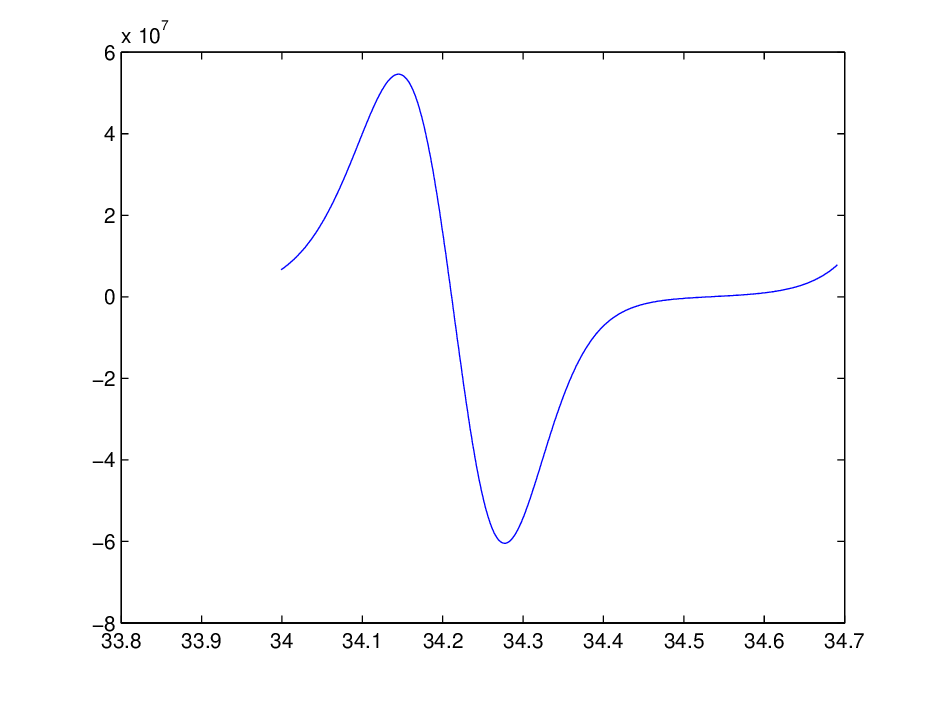}   
\caption{Change of energy rate versus time for a cycle of the Lorenz system as it evolves in one lobe.}
\label{fig:lorenz_rate_energy}
\end{figure}
With a straightforward numerical calculation, we determine that $\frac{dE_K}{dt}$, the rate of change of energy, 
is greater than zero over a cycle. Hence, energy is pumped into the system, Fig:~\ref{fig:lorenz_rate_energy}.
We illustrate in  Appendix B how  the existence of a memory term 
is necessary to have  a positive rate of change of kinetic energy.

\subsection{Potential energy}
\label{subsec:lorenz3}

\begin{figure}[]
\includegraphics[width=8.6cm]{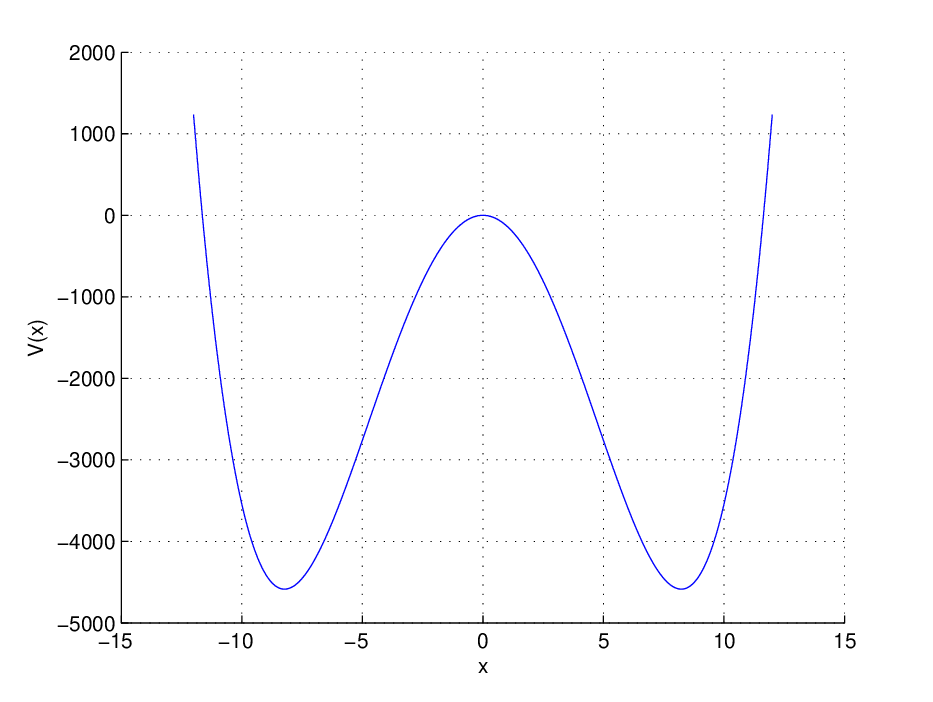}
\caption{Potential of the Lorenz system.}
\label{fig:lorenz_potential}
\end{figure}

As in the R\"ossler system, the presence of an elastic force in the Lorenz IDE is 
suggestive of the existence of a potential.  
The usual
integration of the elastic force, $ U(x)=-\int F dx $,
where $F$ is the force in Eq.~(\ref{eq:lorenz-in-x}),  yields
a quadratic potential featuring two wells 
\begin{equation}
U(q) = -\frac{\sigma(r-1)}{2} q^2 - (1-\frac{b}{2\sigma}) \int q 
\int_{0}^t e^{-b\tau} q^2(t-\tau) d\tau dq + \frac{1}{8} q^4.
\label{eq:lorenz_potential}
\end{equation}
To eliminate the effect of the memory term, 
we replace the kernel of the memory term integral 
with a delta function. 
We used the same approximation in deriving the R\"ossler potential.
In this way, we obtain
an approximate potential as
\begin{equation}
U(q) = -\frac{\sigma(r-1)}{2} q^2 - \frac{1}{4}(1-\frac{b}{2\sigma})q^4 + \frac{1}{8} q^4.
\label{eq:lorenz_potential}
\end{equation}
We plot this potential function in Fig: \ref{fig:lorenz_potential}.

\subsection{The Duffing system and the Lorenz system}
\label{subsec:lorenz4}

There are several commonalities between the Duffing system and the Lorenz system. 
Once the Lorenz equation of
motion is transformed into an equation with one degree of freedom, it is easy to compare the two.
The structure of the second order differential equation of the Duffing system 
is virtually identical to Lorenz's.
In fact, the Duffing system is given by 
\begin{equation}\label{eq:duffing}
\ddot{x} + \gamma \dot{x} - x + x^3 = F cos(\omega t),
\end{equation}
and the Lorenz system is 
\begin{equation} \label{eq:lorenz}
\ddot{x}+(\sigma+1)\dot{x}-\sigma(r-1)x + \frac{1}{2} x^3 = (1-\frac{b}{2\sigma})x \int_{0}^t e^{-b\tau} x^2(t-\tau) d\tau.
\end{equation}
We note that the term $(1-\frac{b}{2\sigma})x \int_{0}^t e^{-b\tau} x^2(t-\tau) d\tau$ is not 
equivalent to the forcing term of the Duffing system $F cos(\omega t)$; the memory term of the 
Lorenz system is not a driving force, it is
a correction term to the elastic force. Therefore,
a more appropriate form of the Duffing equation is obtained by setting 
the force in the Duffing equation to zero, $F=0$; then, 
\begin{equation}\label{eq:duffing2}
\ddot{x} + \gamma \dot{x} - x + x^3 = 0.
\end{equation}
Without the driving term, the oscillating motion of the Duffing system
eventually dies out, as the system is no longer driven by an external force.

We would like to point out the fact that the only difference between the damped 
Duffing equation without driving force and the Lorenz IDE is the
existence of the memory term in the Lorenz system. Therefore,  we 
conjecture that the self-oscillatory character of the Lorenz system
is determined by the influence of its memory term. 

\section{Discussion}
\label{sec:discussion}

The two sets of first order ODEs defining respectively the R\"ossler and Lorenz systems, and their 
equivalent second order IDEs, provide two complementary descriptions of the motion: 
kinematics and dynamics. 
In many ways, the kinematics provides a simpler description as it deals with velocity, position, and time.
In contrast, the dynamics offers a richer description of the motion since it takes into account  
momenta, forces and energy.

The general structure of the IDE of motion for the R\"ossler and for the Lorenz 
systems can be expressed 
in the following form:
\begin{equation} \label{eq:general-form}
\ddot{q} + d(q,\dot{q}) + \int_{t_0}^t e^{-k\tau}
f(q,\dot{q};t-\tau) d\tau + r(q,q^n) +  \int_{t_0}^t e^{-k\tau}
f(q^n;t-\tau) d\tau = F.
\end{equation} 
Here $d(q,\dot{q}) + \int_{t_0}^t e^{-k\tau} f(q,\dot{q};t-\tau) d\tau$ is a frictional term,
$r(q,q^n) +  \int_{t_0}^t e^{-k\tau} f(q^n;t-\tau) d\tau$ is en elastic force, and $F$ is a constant force.
Both, the frictional force and the elastic force show an added term, a convolution integral, 
known in the literature as
heredity, retardation reaction, or  
memory terms \cite{volterra,langevin1}.

Second-order IDEs with
memory terms are commonly found in models of biological, physiological,  financial systems, and, 
more generally, in stochastic systems \cite{biologicalmodels,voit,mechbio}. Also,
dynamical systems specified by second order IDEs are not novel. In fact, 
Volterra derived a
second-order IDE, circa 1930's, to study how
heredity influences  population growth models \cite{brunner1987,volterra}.   This 
equation takes the following form 
\begin{equation}
\ddot{q}(t)=  P_1(t) \dot{q} + \int_{0}^t K_1(t-\tau) \dot{q}(\tau) d\tau + P_0(t) q(t) + \int_{0}^t K_0(t-\tau) q(t) d\tau + f(t), 
\label{eq:volterra_equation} 
\end{equation}
with $P_0$, $P_1$, $f$, $K_0$, and $K_1$ as given continuous functions.
Here, we recognize the frictional term as $P_1(t) \dot{q} + \int_{0}^t K_1(t-\tau) \dot{q}(\tau) d\tau$, the elastic force as
$P_0(t) q(t) + \int_{0}^t K_0(t-\tau) q(t) d\tau$, and the external force as $f(t)$.
We note that the Volterra equation and the R\"ossler and Lorenz IDEs share the same general structure:
the equations represent an oscillator with
frictional and elastic forces having correction terms specified by memory terms \cite{langevin1,langevin2,memory0,memory1}.

We also find memory terms in stochastic system such as the Langevin dynamics that models the dynamics of molecular systems. In particular, we consider the generalized Langevin equation, which
is expressed in terms of a linear IDE
\begin{equation}
m\ddot{q}(t)=-m\int_{-\infty}^t \gamma(t-\tau) \dot{q}(\tau) d\tau + F(t). 
\label{eq:generalized_langevin_equation} 
\end{equation}
that takes into account retardation effects (or memory effects) imposed 
on its frictional force.
The frictional force in Equ:~\eqref{eq:generalized_langevin_equation} is 
determined by a memory term with a characteristic memory kernel $\gamma(t-\tau)$,
which describes the internal inertial properties of the system. 
This convolution integral shows that the velocity of the system does 
not respond instantaneously to the external force $F(t)$ 
but that it depends on the whole history of the process. 
In the same way, the Volterra equation, as well as the R\"ossler and 
Lorenz equations, take into account the entire history of the 
evolution of the systems in their frictional force and elastic force.

The R\"ossler and the Lorenz systems are both dissipative 
and  self-oscillating systems. We tracked down the cause of self-oscillation to the 
ability of the systems to pump energy into their systems. 
In the chaotic regime, over a cycle, the rate of change of 
energy is greater than zero, but
never a constant. On the other hand, limit cycles are
characterized by rate of change of energy equals to zero per each cycle of oscillation.
We conjecture that the chaotic features of these systems are 
linked to the distortions that the memory terms
in the solutions of the equations of motion.

\section{Conclusions}

In this paper we have explicitly derived   a second order 
 IDE of motion for both 
the R\"osssler and the Lorenz systems. In each case we have derived a
potential function and a rate of change of the kinetic energy
over a cycle.   We  found that this rate of change was
greater than zero (0) for the two non-linear systems when in
the chaotic regime, but never a constant,  and  
precisely zero (0) for a limit cycle. 

\begin{appendices}

\section{Appendix A: Three templates of differential equations}

In this appendix, we examine three possible configurations of the equations of motion 
that are determined by modelling classic chaotic attractors as spring-like systems.

We consider the general equation 
\begin{equation} 
\ddot{q}+h(\dot{q},q)+\beta q + \alpha M(q^n) + F = 0.
\end{equation}
The three free parameters that 
define the different types of systems are $\alpha$,  $\beta$  and 
the exponent $n$, $n$ being the degree of correction term 
to the elastic force.
When $\beta$ and $\alpha$ are positive and  $n=3$
  \begin{equation} 
    \begin{cases}{}
      \alpha > 0  \\
      \beta > 0 ;~~~~~~~~~~~\ddot{q}+h(\dot{q},q)+\beta q + \alpha M(q^n) = 0  
    \end{cases}
  \end{equation}
we obtain a model of a hardening spring \cite{diffeq} that evolves in a one well potential 
   (i.e., a potential with one minimum).
This fully describes the Duffing system.

On the other hand, if $\beta$ is negative, $\alpha$ positive, and $n=3$,
  \begin{equation}
    \begin{cases}{}
      \alpha > 0  \\
      \beta < 0 ;~~~~~~~~~~~\ddot{q}+h(\dot{q},q)-\beta q + \alpha M(q^n) = 0  
    \end{cases}
  \end{equation}
we then have a hardening spring in a double well potential (a potential with 
    two minima and one maximum).   This is the case of the Lorenz system.   
If $\beta$ is positive, $\alpha$ negative, and the degree of the correction term is $n=2$,
  \begin{equation}
    \begin{cases}{}
      \alpha < 0  \\
      \beta > 0 ;~~~~~~~~~~~\ddot{q}+h(\dot{q},q)+\beta q - \alpha M(q^2) = 0  
    \end{cases}
  \end{equation}
we obtain a softening spring in one well potential \cite{diffeq}.  
This is the case of the R\"ossler system.
The maxima and minima of these potentials are the fixed points of the 
corresponding attractors.

\section{Appendix B: Memory term as necessary condition for self-oscillation}

For the types of systems discussed in this paper, we have shown that self-oscillatory motion implies positive energy rate of change.
In fact, 
in sections \ref{subsec:rossler3} and \ref{subsec:lorenz2}, we found that during one cycle, the energy rate of change is overall positive when a system
operates in chaotic regime, or zero when a system produces a limit cycle.
However, we note that, during any given cycle, the instantaneous energy rate of change can be positive or negative.
In this appendix we explore the condition that allows the energy rate of change to be positive.
In particular, and specifically for the Lorenz system, 
we show that the existence of a memory term is  
a necessary condition  for self-oscillation.  

Consider the Lorenz IDE, for which we specify 
the control parameters $\sigma = 10$, $b = 8/3$, and $r = 28$,
\begin{equation*}
\ddot{q} + (\sigma+1) \dot{q} - \sigma(r-1)q + 
\frac{1}{2} q^3 + 
\sigma(1-\frac{b}{2\sigma})q \int_{0}^t e^{-b\tau} q^2(t-\tau) d\tau = 0.
\end{equation*}
We rewrite the above equation by multiply the left and right sides of the equation by $\dot{q}$ to
allow to compute the energy rate of change as shown in section  \ref{subsec:lorenz2}.
\begin{equation*}
\ddot{q}\dot{q}=  - (\sigma+1) \dot{q}^2 + \sigma(r-1)q\dot{q} - 
\frac{1}{2} q^3 \dot{q}- 
\sigma(1-\frac{b}{2\sigma})q\dot{q} \int_{0}^t e^{-b\tau} q^2(t-\tau) d\tau.
\end{equation*}
Since $\ddot{q}\dot{q} = \frac{dE_k}{dt}$, the energy rate of change, we focus on the condition that specify $\frac{dE_k}{dt}>0$; therefore, we consider
\begin{equation*}
- (\sigma+1) \dot{q}^2 + \sigma(r-1)q\dot{q} - 
\frac{1}{2} q^3 \dot{q}- 
\sigma(1-\frac{b}{2\sigma})q\dot{q} \int_{0}^t e^{-b\tau} q^2(t-\tau) d\tau > 0.
\end{equation*}
In other words, we have
\begin{equation*}
- (\sigma+1) \dot{q}^2 > - \sigma(r-1)q\dot{q} + 
\frac{1}{2} q^3 \dot{q}+ 
\sigma(1-\frac{b}{2\sigma})q\dot{q} \int_{0}^t e^{-b\tau} q^2(t-\tau) d\tau,
\end{equation*}
or 

\begin{equation*}
 (\sigma+1) \dot{q}^2 <  \sigma(r-1)q\dot{q} - 
\frac{1}{2} q^3 \dot{q}- 
\sigma(1-\frac{b}{2\sigma})q\dot{q} \int_{0}^t e^{-b\tau} q^2(t-\tau) d\tau,
\end{equation*}
Since $\sigma$ is a positive number, $(\sigma+1) \dot{q}^2 > 0$ at all times,
the following
inequaliy must hold: 
\begin{equation*}
 \sigma(r-1)q\dot{q} - 
\frac{1}{2} q^3 \dot{q}- 
\sigma(1-\frac{b}{2\sigma})q\dot{q} \int_{0}^t e^{-b\tau} q^2(t-\tau) d\tau > 0.
\end{equation*}
Factoring $q\dot{q}$ from this inequality yields 
\begin{equation}\label{eq:lorenz_sn_condition1}
 q\dot{q} [\sigma(r-1) - 
\frac{1}{2} q^2 - 
\sigma(1-\frac{b}{2\sigma}) \int_{0}^t e^{-b\tau} q^2(t-\tau) d\tau] > 0.
\end{equation}
The term $q\dot{q}$ can be positive or negative 
(given that $q$ and $\dot{q}$ can be positive or negative at a given time $t$); this therefore leads to 
 Equ:~\ref{eq:lorenz_sn_condition1} being true under two conditions:

\begin{equation*}
q\dot{q} \lessgtr 0~~~\longrightarrow ~~~~[\sigma(r-1) - 
\frac{1}{2} q^2 - 
\sigma(1-\frac{b}{2\sigma}) \int_{0}^t e^{-b\tau} q^2(t-\tau) d\tau] 
\lessgtr 0
\end{equation*}
In turn, these two conditions yield 
\begin{equation*}
\sigma(r-1) - \frac{1}{2} q^2   \lessgtr
\sigma(1-\frac{b}{2\sigma}) \int_{0}^t e^{-b\tau} q^2(t-\tau) d\tau
\end{equation*}
depending on the sign of $q\dot{q}$ at any given $t$.
In either cases, for $\frac{dE_k}{dt} > 0$ to be true, the 
memory term has to exist and
has to be positive, $\int_{0}^t e^{-b\tau} q^2(t-\tau) d\tau > 0$:
\begin{equation*}
\frac{dE_k}{dt} > 0 ~~~\longrightarrow ~~~\int_{0}^t e^{-b\tau} q^2(t-\tau) d\tau > 0.
\end{equation*}
Now, suppose that $\int_{0}^t e^{-b\tau} q^2(t-\tau) d\tau > 0$ 
is not satisfied, for example 
$\int_{0}^t e^{-b\tau} q^2(t-\tau) d\tau = 0$, then 
$\frac{dE_k}{dt} > 0$ would not be true either, with the consequence
\begin{equation*}
\int_{0}^t e^{-b\tau} q^2(t-\tau) d\tau = 0 ~~~\longrightarrow ~~~ \frac{dE_k}{dt} < 0.
\end{equation*}
In fact, without a memory term, equation~\ref{eq:lorenz_sn_condition1} 
would read 
\begin{equation*}
 q\dot{q} [\sigma(r-1) - \frac{1}{2} q^2] < 0 ,
\end{equation*}
corresponding to a pure dissipative system. Therefore,
$\int_{0}^t e^{-b\tau} q^2(t-\tau) d\tau = 0$ implies $\frac{dE_k}{dt} < 0$,
which says that for self-oscillation to occur,
the existence of the memory term 
$\int_{0}^t e^{-b\tau} q^2(t-\tau) d\tau > 0$
is a necessary condition.

\end{appendices}

\bibliography{forces-motion-strange-attractors_pub}
\bibliographystyle{unsrt}
 
\end{document}